\documentclass[11pt]{article}
\usepackage[usenames,dvipsnames,svgnames,table]{xcolor} 

\usepackage{algorithm}
\usepackage{algorithmic}
\floatname{algorithm}{Algorithm}

\usepackage{longtable}
\usepackage{hhline}
\usepackage{float}
\newfloat{algorithm}{tb}{lop}

\usepackage{enumitem} 
\usepackage{rotfloat} 
\usepackage{graphicx}
\usepackage{epsfig}
\usepackage{amssymb, amsmath, amsthm}
\usepackage{verbatim}
\usepackage{natbib}
\usepackage{authblk}
\usepackage{multirow}
\usepackage{subcaption} 
\usepackage{array}
\newcolumntype{L}{>{\centering\arraybackslash}m{0.1\linewidth}}

\newtheorem{theorem}{Theorem}[section]

\DeclareMathOperator*{\argmin}{argmin}
\DeclareMathOperator*{\argmax}{argmax}

\newcommand{\calM}{\mathcal{M}}

\newcommand{\bbR}{\mathbb{R}}
\newcommand{\bbS}{\mathbb{S}}
\newcommand{\bbE}{\mathbb{E}}

\newcommand{\bbX}{\mathbb{X}}

\newcommand{\bfX}{\mathbf{X}}
\newcommand{\bfZ}{\mathbf{Z}}

\newcommand{\bfTheta}{\mathbf{\Theta}}

\newcommand{\Frechet}{Fr\'{e}chet }

\begin{document}
	
\title{Parameter Estimation and Model-Based Clustering with Spherical Normal Distribution on the Unit Hypersphere}
\author[1]{Kisung You}
\affil[1]{Department of ACMS, University of Notre Dame}
\date{}

\maketitle
\begin{abstract}
In directional statistics, the von Mises-Fisher (vMF) distribution is one of the most basic and popular probability distributions for data on the unit hypersphere. Recently, the spherical normal (SN) distribution was proposed as an intrinsic counterpart to the vMF distribution by replacing the standard Euclidean norm with the great-circle distance, which is the shortest path joining two points on the unit sphere. We propose numerical approaches for parameter estimation since there are no analytic formula available. We consider the estimation problems in a general setting where non-negative weights are assigned to observations. This leads to a more interesting contribution for model-based clustering on the unit hypersphere by finite mixture model with SN distributions. We validate efficiency of optimization-based estimation procedures and effectiveness of SN mixture model using simulated and real data examples. 
\end{abstract}


\section{Introduction}\label{sec:intro}

Learning with data on Riemannian manifolds has attained much interests due to its capability that enables us exploit geometric structures of the underlying space behind the data \citep{bhattacharya_nonparametric_2012-1, patrangenaru_nonparametric_2016, pennec_riemannian_2020}. For example, the Stiefel and Grassmann manifolds, which are the spaces of orthonormal bases and subspaces, have been extensively used in computer vision community for applications such as face recognition \citep{aggarwal_system_2004}, shape analysis \citep{goodall_projective_1999}, and human pose modeling \citep{bissacco_recognition_2001}, to name a few. In medical image analysis, the manifold of symmetric and positive definite matrices has been employed for inferential tasks such as modeling anatomical variability in a population of brain scans \citep{fletcher_geometric_2009-1} and translating popular statistical tools for the space of functional brain connectivity \citep{you_re-visiting_2021}. 

In statistics, one of the long-standing disciplines to learn with manifold-valued data is directional statistics  \citep{mardia_directional_2000, ley_modern_2017-1}. As its name suggests, main objects in directional statistics include directions, axes, and rotations where the first two are represented by unit vectors and lines through the origin. Especially, the directions seem to attract the largest portion of attention in directional statistics for its ubiquity as many examples involve observations on the unit hypersphere.

The von Mises-Fisher (vMF) distribution is one of the most well-studied probability distributions on the unit hypersphere that belongs to a location-scale family \citep{fisher_dispersion_1953, mardia_directional_2000}. Given the $p$-dimensional random vector $x$, the probability density is given by
\begin{equation}\label{def:density_vMF}
f_{\text{vMF}}(x|\mu,\kappa) = C_p (\kappa) \exp (\kappa \mu^\top x)
\end{equation}
where $C_p (\kappa)$ is the normalizing constant and $(\mu,\kappa)$ are location and concentration parameters for the distribution. The vMF distribution is an exponential family in that it has played a foundational role in statistical inference for random variables on the unit hypersphere and has been extensively studied from a variety of aspects ranging from parameter estimation \citep{jupp_maximum_1979, sra_short_2012} to finite and nonparametric Bayesian clustering \citep{banerjee_clustering_2005,hornik_movmf_2014, bhattacharya_nonparametric_2012-2}.

\begin{figure}[ht]
	\centering
	\includegraphics[width=0.99\linewidth]{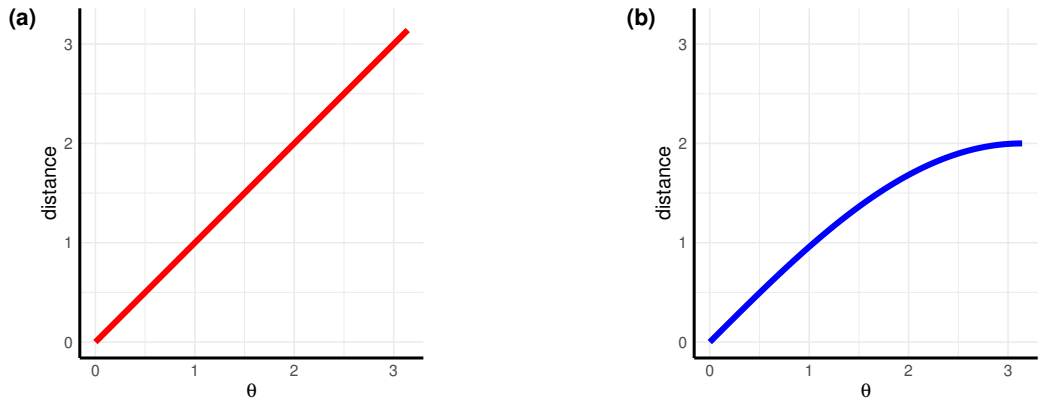}
	\caption{Distances for two points on the great circle displaced by angle $\theta$ under (a) intrinsic and (b) extrinsic frameworks.}
	\label{fig:example_distance}
\end{figure}

It is straightforward to observe that the vMF distribution can be formulated in a similar fashion of the Gaussian distribution as follows,
\begin{equation*}
\exp \left( -\frac{\kappa}{2} \|x-\mu\|^2 \right) = \exp \left( -\frac{\kappa}{2} \left\lbrace x^\top x - 2\mu^\top x + \mu^\top \mu \right\rbrace \right) \propto \exp \left(\kappa \mu^\top x\right)
\end{equation*}
since both $x$ and $\mu$ are unit-norm vectors. In the context of manifold-valued data analysis, the use of Euclidean norm corresponds to the \emph{extrinsic} framework, where a distance between two objects is measured by a standard distance after equivariant embedding of the objects onto the Euclidean space \citep{bhattacharya_nonparametric_2012-1}. A counterpart in the dichotomy of geometric data analysis is the \emph{intrinsic} framework, which measures a distance via the shortest-path geodesic joining two points, which is the great circle on the unit hypersphere. Therefore, replacing the Euclidean norm with the great-circle distance gives a counterpart of the vMF distribution corresponds to the Riemannian normal law proposed in \cite{pennec_intrinsic_2006}. One motivation for such proposal comes from an observation that as two points are farther apart on the hypersphere, the extrinsic distance tends to understate the degree of distantness as shown in Figure \ref{fig:example_distance}.

Recently, the spherical normal (SN) distribution was formally presented whose density is governed by the squared geodesic distance in the spirit of  intrinsic framework \citep{hauberg_directional_2018-1}. While anisotropic and isotropic versions of the SN distribution were originally proposed,  the former has less attractive aspects due to lack finite-time computation of the normalizing constant and complicated optimization routines for parameter estimation. On the other hand, an isotropic version has a closed-form expression of the normalizing constant and has a number of interesting connections to the well-established literature on manifold-valued data analysis. Throughout this work, we limit our focus on the isotropic SN distribution.

When given a proper probability distribution function, one of the first tasks for statistical inference is parameter estimation in the sense of maximum likelihood. Based on an observation that estimating the location parameter of the SN distribution is equivalent to the \Frechet mean estimation problem, we present an algorithm for a more general case where each observation is given a fixed weight independently. The concentration parameter estimation is a more challenging task because of the normalizing constant expressed in an integral form. Similar to the vMF distribution's case, we propose a numerical approach for estimation based on univariate optimization routines and approximation of derivatives via finite difference schemes.

An important application of a probability distribution is model-based clustering \citep{bouveyron_model-based_2019}. The finite mixture model \citep{mclachlan_finite_2019} is one of the central tools in density estimation and model-based clustering by matching each observation to the cluster of maximal probability. Based on the aforementioned routines, we present the finite mixture of SN distributions where each component is the SN distribution, similar to the mixture of vMF distributions  \citep{banerjee_clustering_2005, hornik_movmf_2014}.

The rest of the paper is organized as follows. Section \ref{sec:preliminaries} reviews the elements of Riemannian geometry on the unit hypersphere and the SN distribution. Section \ref{sec:estimation} presents theoretical results regarding existence and uniqueness of the maximum likelihood estimates and numerical optimization routines for estimation. In Section \ref{sec:clustering}, We present computational routines of the SN mixture for clustering with some variants. We validate the algorithms with simulated and real data examples in Section \ref{sec:experiment}. We conclude in Section \ref{sec:conclusion} by discussing issues and directions for future studies.

\section{Preliminaries}\label{sec:preliminaries}

We start this section by introducing notations. Let $\bbS^p = \lbrace x \in \bbR^{p+1} : \|x\|_2 = 1 \rbrace $ denote a $p$-dimensional sphere in a $(p+1)$-dimensional Euclidean space $\bbR^{p+1}$. For a general metric space $(\bbX, d)$, we denote the open ball of radius $r > 0$ centered at a point $x \in \bbX$ as $B (x, \epsilon) := \lbrace y\in\bbX ~|~ d(x,y) < \epsilon \rbrace $. $\bbR^+$ stands for a set of strictly positive real numbers. For $x,y\in \bbS^p$, the distance $d(x,y)$ refers to the geodesic distance on the sphere while $\|x-y\|$ is the standard $L_2$ distance in an ambient Euclidean space $\bbR^{p+1}$. For the rest of this paper, several terms such as sphere, unit sphere, and unit hypersphere may be used interchangeably.

\subsection{Elements of Riemannian Geometry on the Sphere}

We first review some properties of the unit sphere as a Riemannian manifold along with explicit expressions of operations for computation \citep{absil_optimization_2008-1}. For an introduction and general exposition to Riemannian manifolds in detail, we refer interested readers to some standard references \citep{carmo_riemannian_1992, lee_riemannian_1997}.

\begin{figure}[ht]
	\centering
	\includegraphics[width=.5\linewidth]{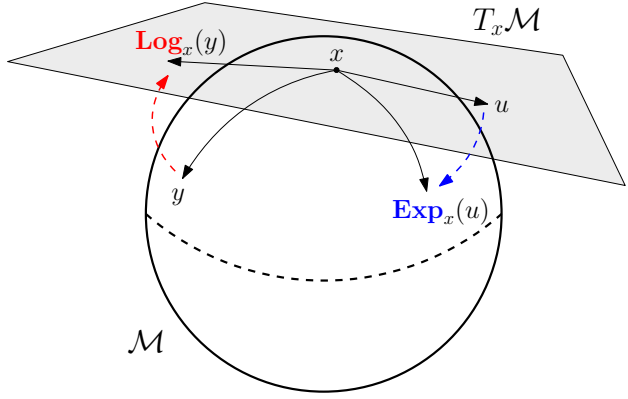}
	\caption{Exponential and logarithmic maps on a Riemannian manifold $\calM$.}
	\label{fig:maps}
\end{figure}

The tangent space at $x \in \bbS^p$ is given by $T_x \bbS^p = \lbrace y \in \bbR^{p+1}:\langle x, y \rangle = 0\rbrace$ where $\langle \cdot, \cdot \rangle$ is a standard inner product. 
As a Riemannian manifold, $\calM = \bbS^p$ is equipped with a canonical metric $g_x (u,v) = \langle u, v \rangle$ for any $u,v \in T_x \bbS^p$. The unit sphere is known to have a positive constant sectional curvature of 1. The shortest path connecting two points $x,y \in \bbS^p$ is the great circle so that the distance between $x$ and $y$ is given by $d(x,y) = \cos^{-1}(\langle x, y \rangle)$. An exponential map is a map from a tangent space to a manifold itself and its inverse is called a logarithmic map, which are shown in Figure \ref{fig:maps}. For $x,y \in \bbS^p$ and $u \in T_x \bbS^p$, two maps are given as follows,
\begin{subequations}\label{eq:exp_log_maps}
\begin{align}
\text{Exp}_x (u) &= \cos (\|u\|) x + \frac{\sin(\|u\|)}{\|u\|} u \\
\text{Log}_x (y) &= \frac{d(x,y)}{\|\text{Proj}_x (y-x)\|} \text{Proj}_x (y-x)
\end{align}	
\end{subequations}
where $\text{Proj}_x (z) = z - \langle x, z \rangle x$ is a projection operator for a vector $z \in \bbR^{p+1}$ onto $T_x \bbS^p$. Finally, the injectivity radius, which is a maximal radius where the exponential map is a diffeomorphism, for the unit sphere is $\pi$. 

\subsection{Spherical Normal Distribution}

\begin{figure}[ht]
	\centering
	\includegraphics[width=0.9\linewidth]{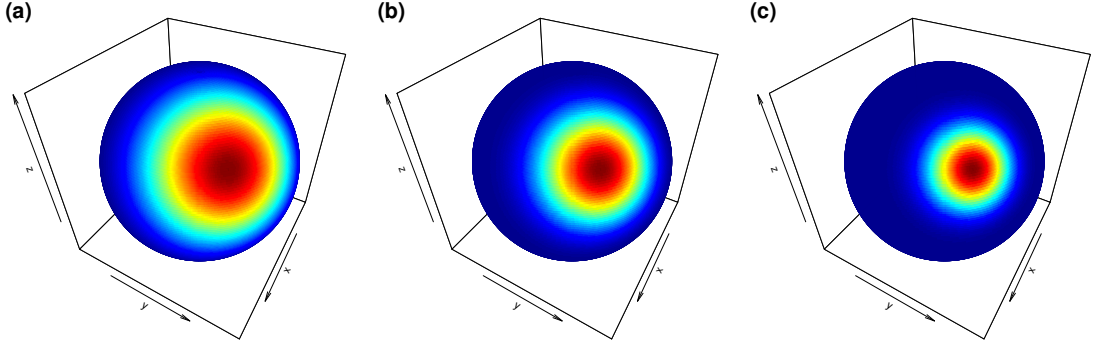}
	\caption{Densities of the isotropic SN distributions on $\bbS^2$ with different concentration parameter values of (a) $\lambda=5$, (b) $\lambda=10$, and (c) $\lambda=20$}
	\label{fig:example_density3}
\end{figure}

As introduced before, we limit our focus on an isotropic version of the SN distribution that was formally presented in \cite{hauberg_directional_2018-1}. Given location and concentration parameters $\mu \in \bbS^p$ and $\lambda \in \bbR^+$, the density function of the SN distribution is given by 
\begin{equation}\label{def:density_SN}
f_{\textsf{SN}}(x|\mu,\lambda) = \frac{1}{Z_p(\lambda)} \exp \left(-\frac{\lambda}{2}d^2 (x,\mu)\right)
\end{equation}
where the normalizing constant $Z_p (\lambda)$ is given in an integral from by 
\begin{equation}\label{def:normalizing_constant}
\begin{split}
Z_p (\lambda) &= \int_{\bbS^p} \exp \left(-\frac{\lambda}{2}d^2 (x,\mu)\right) dx \\
&= A_{p-1} \int_{r=0}^{\pi} \exp\left(-\frac{\lambda r^2}{2}\right) \sin^{p-1}(r)dr
\end{split}
\end{equation}
where $A_{p-1} = 2\pi^{p/2}/\Gamma(p/2)$ is the hypervolume or surface area of $\bbS^{p-1}$ and $\Gamma(\cdot)$ is the standard Gamma function. Roughly speaking, the concentration parameter $\lambda$ is an inverse of variance in an epistemological sense as shown in Figure \ref{fig:example_density3} where a large $\lambda$ leads to concentrated mass at the vicinity of $\mu$ and a small $\lambda$ shows dispersed distribution of the mass over a larger support. 

\section{Maximum Likelihood Estimation}\label{sec:estimation}

Let $\bfX = \lbrace x_1, \ldots, x_n\rbrace$ a random sample on the unit sphere $\bbS^p$. The maximum likelihood estimates $(\hat{\mu}_{\text{MLE}}, \hat{\lambda}_{\text{MLE}}) \in \Theta = \bbS^p \times \bbR^+$ are obtained by maximizing the following log-likelihood function
\begin{equation}\label{def:lkd}
\begin{split}
L(\mu,\lambda; \bfX) &= \sum_{i=1}^n \log \left\lbrace \frac{1}{Z_p(\lambda)} \exp \left(-\frac{\lambda}{2}d^2(x_i,\mu)\right)\right\rbrace\\ 
&= -\frac{\lambda}{2}\sum_{i=1}^n d^2 (x_n,\mu) - n\log Z_p(\lambda)
\end{split}
\end{equation}
which does not admit closed-form solutions due to the existence of nonlinear terms in a likelihood function hence necessitates to employ numerical optimization. To briefly describe the strategy for both theory and computation, we recognize that the constrained maximization of Equation \eqref{def:lkd} with respect to $\mu$ does not depend on the concentration parameter,
\begin{equation}\label{def:problem_mu}
\hat{\mu}_{\text{MLE}} = \underset{\mu \in \bbS^p}{\argmax} -\frac{\lambda}{2} \sum_{i=1}^n d^2(x_i,\mu) = \underset{\mu \in \bbS^p}{\argmin} \sum_{i=1}^n  d^2 (x_i, \mu)
\end{equation}
where the above equation has long been known as  \Frechet or Karcher mean problem \citep{frechet_les_1948, grove_how_1973-1, afsari_riemannian_2011}.  Given $\hat{\mu}$, solving for $\hat{\lambda}$ reduces to the following optimization problem
\begin{equation}\label{def:problem_lambda}
\hat{\lambda}_{\text{MLE}} = \underset{\lambda \in \bbR^+}{\argmax} -\frac{\lambda}{2} \sum_{i=1}^n d^2(x_i,\hat{\mu}) - n \log Z_p(\lambda) =  \underset{\lambda \in \bbR^+}{\argmin} ~\hat{C} \lambda + \log Z_p (\lambda)
\end{equation}
for a known positive constant $\hat{C} = \sum_{i=1}^n d^2(x_i,\hat{\mu}_{\text{MLE}})/2n$. The following theorem shows existence and uniqueness of maximum likelihood estimates.

\begin{theorem}\label{thm:hadamard}
	Let $x_1,x_2,\ldots,x_n$ be a random sample on $\bbS^p$ that are contained in an open geodesic ball $B(x, \pi/2)$ for some $x \in \bbS^p$. Then, maximum likelihood estimates of $(\mu,\lambda)$ uniquely exist. 
\end{theorem} 
\begin{proof}
	The existence and uniqueness of $\hat{\mu}$ has been well-studied on a general Riemannian manifold $\calM$ \citep{kendall_probability_1990-1, nielsen_medians_2013}. Let $r^* = \min \lbrace \text{inj}(\calM), \pi/\sqrt{\bar{C}} \rbrace$ where $\bar{C}$ is the least upper bound of sectional curvatures of $\calM$.  Proposition 5.2 of \cite{bhattacharya_nonparametric_2012-1} states that if a measure $Q= (1/n) \sum_{i=1}^n \delta_{x_i}$ on $\calM$ has support in $B(x, r^*/2)$ for some $x \in \bbS^p$, then $Q$ has a unique local sample mean inside the ball. On the sphere $\calM = \bbS^p$, twice the convexity radius is $r^* = \min \lbrace \pi, \pi/1 \rbrace = \pi$, which proves the existence and uniqueness of $\hat{\mu}_{\text{MLE}}$.
	
	For $\hat{\lambda}_{\text{MLE}}$, we first show that there exists a unique critical point of the Equation \eqref{def:problem_lambda}, which specifies the likelihood in an equivalent form. Let $g(\lambda) = \hat{C} \lambda + \log Z_p (\lambda)$ with a known constant $\hat{C} = \sum_{i=1}^n d^2 (x_i, \hat{\mu}_{\text{MLE}}) / 2n$ where $\hat{C} \in \lbrack 0, \frac{\pi^2}{2}\rbrack$ since the geodesic distance between any two points on the unit sphere is upper bounded by the injectivity radius $\pi$. Since $g'(\lambda) = 0$ if and only if $\hat{C} Z_p (\lambda) + Z_p ' (\lambda) = 0$, it suffices to show that $T(C) = C Z_p (\lambda) + Z_p '(\lambda)$ has a unique solution in $( 0, \frac{\pi^2}{2} )$ for any given $\lambda$. This implies a bijection between $\lambda$ and $C$ upon which the existence and uniqueness of a critical point for $g(\lambda)$ is established. First, we have $T(0) < 0$ since 
	\begin{equation}\label{eq:auxiliary}
	T(C) = A_{p-1} \int_{r=0}^\pi \left(C - \frac{r^2}{2}\right) \exp\left(-\frac{\lambda r^2}{2}\right) \sin^{p-1}(r)dr
	\end{equation}
	and $T(\frac{\pi^2}{2}) > 0$ as $C - \frac{r^2}{2} = \frac{\pi^2 - r^2}{2} >= 0$ on the domain. Second, it is trivial that $T(C)$ is continuous as the integrand consists of smooth and bounded functions on a finite interval. Furthermore, $T(C)$ is a monotonically increasing function since for any $C' > C$,
	\begin{equation*}
	T(C')-T(C) = A_{p-1} \int_{r=0}^\pi \left(C'-C\right) \exp\left(-\frac{\lambda r^2}{2}\right) \sin^{p-1}(r)dr > 0.
	\end{equation*}
	Therefore, by the intermediate value theorem and  monotonicity, $T(C)$ has a unique zero in $( 0, \frac{\pi^2}{2} )$ and so does $g'(\lambda)$.
	
	Let $\lambda^*$ be a unique critical point of $g(\lambda)$ and the rest is to show whether $\lambda^*$ is a local minimum. This is attained by comparing $g(\lambda^*+\epsilon)$ and $g(\lambda^* - \epsilon)$ with $g(\lambda^*)$ for $ \forall \epsilon > 0$. First, 
	\begin{align*}
	g(\lambda^* + \epsilon) &= \hat{C} (\lambda^* + \epsilon) + \log Z_p (\lambda^* + \epsilon) > \hat{C} \lambda^* + \hat{C} \epsilon + \log Z_p(\lambda^*) - \frac{\pi^2 \epsilon}{2} = g(\lambda^*) + \epsilon \left(\hat{C} - \frac{\pi^2}{2}\right)
	\end{align*}
	where $\hat{C} - \pi^2/2 \leq 0$. Similary, we have
	\begin{equation*}
	g(\lambda^*-\epsilon) = \hat{C}(\lambda^*-\epsilon) + \log Z_p (\lambda^* - \epsilon) > \hat{C}\lambda^* + \log Z_p (\lambda^*) - C\epsilon = g(\lambda^*) - \hat{C}\epsilon.
	\end{equation*}
	Since both inequalities hold for arbitrarily small $\epsilon$, taking the limit inferior as $\epsilon \rightarrow 0$ gives that both $g(\lambda^*+\epsilon)$ and $g(\lambda^*-\epsilon)$ are greater than $g(\lambda^*)$. Therefore, $g(\lambda)$ attains the minimum at $\lambda = \lambda^*$. 
\end{proof}

We note that the support condition $B(x,\pi/2)$ is natural especially when the manifold of interest is the unit sphere. If we consider north and south poles $\bbS^2$, there exist an infinite number of points along the equator that minimize the sum of squared distances.

\subsection{Estimating the Location}

It was observed from Equation \eqref{def:problem_mu} that  maximum likelihood estimation of $\mu$ is equivalent to the \Frechet mean problem that minimizes sum of squared distances. We illustrate a more general case of weighted \Frechet mean problem for future use in fitting a mixture model. Given a random sample $ \bfX = \lbrace x_1, \ldots, x_n \rbrace$ and non-negative weights $\mathbf{w} = \lbrace w_1, \ldots, w_n\rbrace$, the weighted \Frechet mean problem is given by
\begin{equation}\label{def:problem_weighted_frechet}
\underset{\mu\in\bbS^p}{\min}~f(\mu| \bfX, \mathbf{w}) = \underset{\mu\in\bbS^p}{\min}~ \sum_{i=1}^n w_i d^2 (x_i, \mu)
\end{equation}
hence the maximum likelihood estimate $\hat{\mu}_{\text{MLE}}$ is a solution to the special case of Equation \eqref{def:problem_weighted_frechet} when all weights $w_i$'s are set equal such as $w_i = 1/n$ from an empirical measure point of view. 

The hands-on choice in optimization on manifolds is Riemannian gradient descent \citep{absil_optimization_2008-1}. The gradient descent algorithm for a function $f (x) :\bbR^d \rightarrow \bbR$ evolves iterates along a path characterized by local gradient evaluations, i.e., $x^{(t+1)} = x^{(t)} - \alpha \nabla f (x^{(t)}) $ for some step-size $\alpha$. For a scalar-valued function on a Riemannian manifold $\calM$, the gradient is formerly defined as a vector field. This means for each point $p \in \calM$, $\text{grad} f \rvert_p$ is a tangent vector in $T_p \calM$ so that the resulting update rule requires an additional step to push at iterate tangent vector onto $\calM$ using an exponential map. Since $\text{grad} ~d^2 (x,y) = -2 \textrm{Log}_x y$ for $x,y \in \calM$ which can be easily shown using the variation of energy, we can write the gradient of Equation \eqref{def:problem_weighted_frechet} at iteration $t$ as 
\begin{equation*}
\text{grad} f \rvert_{\mu^{(t)}} = -2 \sum_{i=1}^n w_i \text{Log}_{\mu^{(t)}} (x_i)
\end{equation*}
and the final updating rule is given by 
\begin{equation*}
\mu^{(t+1)} \leftarrow \text{Exp}_{\mu^{(t)}} \left( 2\alpha^{(t)}  \sum_{i=1}^n w_i \text{Log}_{\mu^{(t)}} (x_i) \right)
\end{equation*}
for an initial point $\mu^{(0)}$ and step-size $\alpha^{(t)}$. Explicit forms for exponential and logarithmic maps are given in Equations \eqref{eq:exp_log_maps} and the procedure for location estimation is summarized in Algorithm \ref{code:lkd_mean}. 

\begin{algorithm}
	\caption{Weighted \Frechet mean computation}
	\label{code:lkd_mean}
	\begin{algorithmic}
		\REQUIRE a random sample $\lbrace x_1,\ldots,x_n\rbrace \subset \bbS^p$, weights $\lbrace w_1, \ldots, w_n\rbrace$, stopping criterion $\epsilon$.
		\ENSURE $\hat{\mu} = \argmin f(\mu)$ where $f(\mu) =  \sum_{i=1}^n w_i d^2 (x_i, \mu)$ for $\mu \in \bbS^p$. 
		\STATE Initialize $\mu^{(0)} = \sum_{i=1}^n w_i x_i / \| \sum_{i=1}^n w_i x_i\|$.
		\REPEAT 
			\STATE $\nabla f \vert_{\mu^{(t)}} \leftarrow -2 \sum_{i=1}^n w_i \text{Log}_{\mu^{(t)}} (x_i)$ 
			\STATE $\mu^{(t+1)} \leftarrow \text{Exp}_{\mu^{(t)}} \left( -\alpha^{(t)} \nabla f \vert_{\mu^{(t)}} \right)$
		\UNTIL $\|\nabla f \vert_{\mu^{(t)}} \| < \epsilon$ or $\| \mu^{(t)} - \mu^{(t+1)} \| < \epsilon$ .
	\end{algorithmic}
\end{algorithm}

We illustrate components of the Algorithm \ref{code:lkd_mean}. First, we use an initial starting point $\mu^{(0)} = \sum_{i=1}^n w_i x_i / \| \sum_{i=1}^n w_i x_i\|$. When $w_i$'s are all equal, this corresponds to an extrinsic mean \citep{bhattacharya_nonparametric_2012-1} or an estimate for a location parameter of the vMF distribution. When the support condition of Theorem \ref{thm:hadamard} is satisfied, estimates of both extrinsic and intrinsic means are contained in a convex geodesic ball. In addition to computational benefit that obtaining an extrinsic mean involves addition of vectors and $L_2$ normalization, it is a reasonable choice of starting point to have a small number of iterations for convergence. Second, the step-size parameter $\alpha^{(t)}$ can be determined by line search or simply set as a fixed scalar. The latter strategy was used in Algorithm 1 of \cite{hauberg_directional_2018-1} with $\alpha=0.25$. This approach has trade-offs in the sense that a fixed step-size may incur additional iterations while it avoids repetitive evaluations of the objective function during a single line search. Finally, the stopping criterion $\|\nabla f \vert_{\mu^{(t)}} \| < \epsilon$ is motivated from the fact that $d^2(x,y) = \| \text{Log}_x y \|^2$ for $x,y\in \bbS^p$ with the canonical metric. Also $\| \mu^{(t)} - \mu^{(t+1)}\|$ is justified from the fact that for a Riemannian manifold $\calM$ embedded in Euclidean space, intrinsic and extrinsic distances converge as two points get sufficiently close, i.e.,  $\lim_{x\rightarrow y} d (x,y)/\|x-y\| =1$.

\subsection{Estimating the Concentration}

The maximum likelihood estimation for concentration parameter $\lambda$ is equivalent to solve the following optimization problem
\begin{equation*}
\hat{\lambda}_{\text{MLE}} =\underset{\lambda \in \bbR^+}{\argmin} ~ g(\lambda) \text{ where } g(\lambda) =\hat{C} \lambda + \log Z_p (\lambda)\text{ and } \hat{C} = \frac{1}{2n} \sum_{i=1}^n d^2(x_i, \hat{\mu}_{\text{MLE}})
\end{equation*}
Theorem \ref{thm:hadamard} showed that the problem \eqref{def:problem_lambda} has a unique critical point so that the optimization with respect to $\lambda$ can be cast as a  root-finding problem of $g'(\lambda)=0$. 

Householder's method is a class of univariate root-finding algorithms when the target function is $d$ times continuously differentiable \citep{householder_numerical_1970}. For solving $f(x)=0$ where the function has continuous derivatives up to order $d$, the method iterates by
\begin{equation}\label{algorithm:householder}
x^{(t+1)} \leftarrow  x^{(t)} + d \frac{(1/f)^{(d-1)} (x^{(t)})}{(1/f)^{(d)} (x^{(t)})}
\end{equation}
starting from an initial point $x^{(0)}$. For low orders $d=1$ and $2$, the updating rule in Equation \eqref{algorithm:householder} translates to those found from the Newton's and Halley's method, respectively. Therefore, we derive the updating rules to solve $g'(\lambda)=0$ using Householder's methods of orders 1 and 2 as follows,
\begin{subequations}\label{algorithm:householder_ordered}
\begin{align}
(\textrm{exact Newton's}) \quad&
\lambda^{(t+1)} \leftarrow  \lambda^{(t)} - \frac{g' (\lambda^{(t)})}{g'' (\lambda^{(t)})}  \label{algorithm:householder_newton}\\
(\textrm{exact Halley's}) \quad & \lambda^{(t+1)} \leftarrow \lambda^{(t)} - \frac{2g'(\lambda^{(t)}) g''(\lambda^{(t)})}{2 g'' (\lambda^{(t)})^2 - g'(\lambda^{(t)}) g'''(\lambda^{(t)})}  \label{algorithm:householder_halley}
\end{align}
\end{subequations}
beginning with an initial guess $\lambda^{(0)} \in \bbR^+$. 

Both updating rules involve evaluation of derivatives up to order 2 or 3 at each iteration, which may incur numerical instabilities due to computational reasons such as integrating high-order terms and others. 
In order to make computation more robust, we  employ the centered finite difference schemes using the five-point utensil 
\begin{subequations}\label{algorithm:utensil}
	\begin{align}
	g'(\lambda) &= \frac{g(\lambda+h) - g(\lambda-h)}{2h} =: \frac{A}{2h} \\
	g''(\lambda) &= \frac{g(\lambda+h)-2g(\lambda)+g(\lambda-h)}{h^2} =: \frac{B}{h^2} \\
	g'''(\lambda) &= \frac{g(\lambda+2h)-2g(\lambda+h)+2g(\lambda-h)-g(\lambda-2h)}{2h^3} =: \frac{C}{2h^3}
	\end{align}
\end{subequations}
for a sufficiently small step-size $h>0$ and the derivatives are approximated by a term of order $h^2$ \citep{burden_numerical_2016}. By plugging in the approximate derivatives \eqref{algorithm:utensil} to the updating rules in Equations \eqref{algorithm:householder_ordered}, we get
\begin{align}
(\textrm{approximate Newton's}) \quad&
\lambda^{(t+1)} \leftarrow  \lambda^{(t)} - \frac{h}{2} \left(\frac{A^{(t)}}{B^{(t)}}\right) \label{algorithm:final_newton}\\
(\textrm{approximate Halley's}) \quad & \lambda^{(t+1)} \leftarrow \lambda^{(t)} - 4h \left(\frac{
	A^{(t)}B^{(t)}
}{
8 (B^{(t)})^2 - A^{(t)}C^{(t)}
}\right)  \label{algorithm:final_halley}
\end{align}
where $A^{(t)} = g(\lambda^{(t)}+h) - g(\lambda^{(t)}-h)$ and others defined similarly. 

\begin{algorithm}
	\caption{Maximum likelihood estimation of the concentration parameter}
	\label{code:lkd_lambda}
	\begin{algorithmic}
		\REQUIRE a random sample $\lbrace x_1,\ldots,x_n\rbrace \subset \bbS^p$, a constant $C$, step-size $h$, stopping criterion $\epsilon$.
		\ENSURE $\hat{\lambda} = \argmin g(\lambda)$ where $g(\lambda) = C \lambda + \log Z_p (\lambda)$		
		\STATE Initialize $\lambda^{(0)}$.
		\REPEAT 
		\STATE Update $\lambda^{(t+1)}$ by either Newton's  \eqref{algorithm:final_newton} or Halley's \eqref{algorithm:final_halley} method.
		\UNTIL $|\lambda^{(t)} - \lambda^{(t+1)}|<\epsilon$.
	\end{algorithmic}
\end{algorithm}

We close this section by noting that computational complexities of the exact and approximate methods are compatible. At each iteration, the approximate Newton's method shown in Equation \eqref{algorithm:final_newton} requires 3 integral evaluations, which is same as that of the exact Newton's method \eqref{algorithm:householder_newton}. While the exact Halley's method \eqref{algorithm:householder_halley} evaluates $Z_p^{(k)} (\lambda)$ for $k = 0,\ldots,3$, the approximate Halley's algorithm requires an additional integral computation to evaluates five points around $\lambda^{(t)}$. However, the added computational cost is very thin since the integrand well behaves and the domain of integration is small.

\section{Model-Based Clustering}\label{sec:clustering}

One of the most popular probabilistic clustering methods is finite mixture models \citep{mclachlan_finite_2019}. The density of a finite mixture model with $K$ SN components is given by
\begin{equation*}
h(x|\bfTheta) = \sum_{k=1}^K \pi_k f_{\textsf{SN}}(x|\mu_k,\lambda_k)
\end{equation*}
where $\pi_k \in [0,1]$ and $\sum_{k=1}^K \pi_k = 1$ with parameters $\bfTheta = \lbrace \pi_k, \mu_k, \lambda_k\rbrace_{k=1}^K$. For a random sample $\bfX = \lbrace x_1,x_2,\ldots,x_N\rbrace$ on the unit sphere, the log-likelihood is written as 
\begin{equation}\label{clustering:loglkd_naive}
\log P(\bfX|\bfTheta) = \sum_{n=1}^N \log h(x_n|\bfTheta) = \sum_{n=1}^N \log \left( \sum_{k=1}^K \pi_k f_{\textsf{SN}}(x_n|\mu_k,\lambda_k) \right).
\end{equation} The standard approach to maximize Equation \eqref{clustering:loglkd_naive} is to use Expectation-Maximization (EM) algorithm \citep{dempster_maximum_1977} by introducing latent variables for class membership. We leave detailed description of the technique to standard references in machine learning \citep{bishop_pattern_2006, hastie_elements_2009-1}. Let $\bfZ \in \lbrace 0, 1 \rbrace^{N\times K}$ be a 0-1 matrix of class memberships such that  $\sum_{k=1}^K z_{nk} = 1$ for all $n=1,\ldots,N$. Then, a joint distribution of $\bfX$ and $\bfZ$ is 
\begin{equation}\label{clustering:joint_density}
P(\bfX, \bfZ | \bfTheta) = \prod_{n=1}^{N} \prod_{k=1}^{K} \left\lbrace \pi_k f_{\textsf{SN}}(x_n\vert \mu_k, \lambda_k) \right\rbrace^{z_{nk}}
\end{equation}
Given an initial setting for the parameters $\bfTheta^{(0)}$, EM algorithm alternates the following two steps. In the E-step, the posterior distribution $P(\bfZ\vert \bfX, \bfTheta^{(t)})$ is evaluated to compute the complete-data log likelihood $Q(\bfTheta; \bfTheta^{(t)}) = \bbE_{\bfZ \vert \bfX, \bfTheta^{(t)}} \lbrack \log P(\bfX, \bfZ \vert \bfTheta)\rbrack = \sum_{\bfZ} P (\bfZ \vert \bfX, \bfTheta^{(t)}) \log P(\bfX, \bfZ \vert \bfTheta)$. The M-step attains a new iterate $\bfTheta^{(t+1)}$ by maximizing $Q(\bfTheta; \bfTheta^{(t)})$.

We first present a standard EM algorithm and elaborate on explicit expressions for updating rules. At iteration $t$, the E-step reduces to evaluating $\bbE[z_{nk}]$ that has a universal form for mixture models,
\begin{equation}\label{clustering:Estep_soft}
\gamma_{nk} := \bbE[z_{nk}] = \frac{\pi_k^{(t)} f_{\textsf{SN}} (x_n\vert \mu_k^{(t)}, \lambda_k^{(t)})}{\sum_{j=1}^K \pi_j^{(t)} f_{\textsf{SN}} (x_n\vert \mu_j^{(t)}, \lambda_j^{(t)})}
\end{equation}
for $n=1,\ldots,N$ and $k=1,\ldots,K$. The matrix $\Gamma := \gamma_{nk} \in [0,1]^{N\times K}$ is called a soft clustering or membership matrix which encodes the probability of an observation $x_n$ belonging to the $k$-th cluster. This leads to the complete-data log likelihood at iteration $t$,
\begin{equation}\label{clustering:rule_M_equation}
Q(\bfTheta; \bfTheta^{(t)}) = \sum_{n=1}^N \sum_{k=1}^K \gamma_{nk} \left\lbrace \log \pi_k + \log f_{\textsf{SN}} (x_n\vert \mu_k, \lambda_k) \right\rbrace.
\end{equation}

The M-step maximizes Equation \eqref{clustering:rule_M_equation} by the following update rules. The weight parameters $\pi_k$'s are updated by
\begin{equation}\label{clustering:rule_M_weight}
\pi_k^{{(t+1)}} = \frac{\sum_{n=1}^N \gamma_{nk}}{N}\quad\text{  for }~k=1,\ldots,K
\end{equation}
using the method of Lagrange multipliers to handle the equality constraint $\sum_{k=1}^K \pi_k = 1$. 
Other parameters do not admit closed-form expressions in that they are obtained by numerical optimization introduced in Section \ref{sec:estimation}. The location parameters $\mu_k$'s are solutions of weighted \Frechet mean problem 
\begin{equation}\label{clustering:rule_M_location}
\mu_k^{(t+1)} = \underset{\mu \in \bbS^p}{\argmin} \sum_{n=1}^N  \gamma_{nk} \cdot d^2(x_n, \mu) 
\end{equation}
which can be solved by Algorithm \ref{code:lkd_mean} with weights $w_i = \gamma_{ik},~i=1,\ldots,n$. Similarly, the concentration parameters are solutions of the following problems,
\begin{subequations}\label{clustering:rule_M_concentration}
	\begin{equation}
	\lambda_k^{(t+1)} = \underset{\lambda \in \bbR^+}{\argmin}~
	\left( \frac{\sum_{n=1}^N d^2(\mu_k^{(t+1)}, x_n) \cdot \gamma_{nk}}{ 2 \sum_{n=1}^N \gamma_{nk}} \right) \lambda + \log Z_p (\lambda) \label{clustering:rule_M_concentration_hetero}
	\end{equation}
	for $k=1,\ldots,K$. When all concentration parameters $\lambda_1,\ldots,\lambda_K$ are required to be equal, the model is called homogeneous and the update rule for a common $\lambda$ is as follows,
	\begin{equation}
	\lambda^{(t+1)} =\underset{\lambda \in \bbR^+}{\argmin}~ \left( \frac{1}{2N}\sum_{n=1}^N \sum_{k=1}^K d^2 (\mu_k^{(t+1)}, x_n)\cdot\gamma_{nk} \right)\lambda +  \log Z_p(\lambda). \label{clustering:rule_M_concentration_homo}
	\end{equation}
\end{subequations}
where for both cases the update can be computed by Algorithm \ref{code:lkd_lambda} with the constant term changing along the iterations.

From Equations \eqref{clustering:rule_M_location} and \eqref{clustering:rule_M_concentration}, we can observe that the large number of observations $N$ intensifies computational costs for parameter estimation. We present hard and stochastic assignment heuristics that manipulate a clustering membership matrix $\Gamma$ that represents distribution of the hidden variables. Let the $n$-th row of $\Gamma$ be denoted as $\Gamma_n = [\gamma_{n1},\ldots,\gamma_{nK}]$. The hard assignment is given by
\begin{equation}\label{clustering:Estep_hard}
\textsf{hard}(\gamma_{nk}) = \left. 
\begin{cases}
1, & \text{if $k$ is an index for maximum of $\Gamma_n$,}\\
0, & \text{otherwise}
\end{cases}
\right.
\end{equation}
and the stochastic assignment is similarly written as 
\begin{equation}\label{clustering:Estep_stochastic}
\textsf{stochastic}(\gamma_{nk}) = \left.
\begin{cases}
1, & \text{if $k$ = \textsf{sample}($1:K$, probability=$\Gamma_n)$}\\
0, & \text{otherwise}
\end{cases}
\right.
\end{equation}
where the \textsf{sample} procedure draws a random integer-valued index from 1 to K with probability $\Gamma_n$. Both heuristics aim at making $\Gamma$ sparse. Given a large number of observations, this helps significantly reduce intermediate evaluations and space complexity in update rules for both concentration and location parameters. Furthermore, hard assignment has proven to be optimal in the sense that the scheme maximizes a lower bound on the incomplete-data log likelihood \citep{banerjee_clustering_2005}. The complete procedure is summarized in Algorithm \ref{code:em_spnorm}.

\begin{algorithm}[!t]
	\caption{EM algorithm for mixture of spherical normal distributions.}
	\label{code:em_spnorm}
	\begin{algorithmic}
		\REQUIRE a random sample $\lbrace x_1,\ldots,x_N\rbrace \subset \bbS^p$, number of clusters $K$.
		\ENSURE a clustering membership matrix $\Gamma$.		
		\STATE Initialize $ \bfTheta^{(0)} = \lbrace \pi_k, \mu_k, \lambda_k\rbrace_{k=1}^K$.
		\REPEAT 
			\STATE \{E-step\}
			\FOR{$n=1:N$}
				\FOR {$k=1:K$ }
				\STATE $\Gamma (n,k)= \pi_k^{(t)} f_{\textsf{SN}} (x_n\vert \mu_k^{(t)}, \lambda_k^{(t)})$
				\ENDFOR
				\STATE $\Gamma(n,:) = \Gamma(n,:)/ \sum_{k=1}^K \Gamma(n,k)$
			\ENDFOR 
			\STATE \{Heuristics\}
			\IF {hard assignment}
				\STATE $\Gamma \leftarrow \textsf{hard}(\Gamma)$ by Equation \eqref{clustering:Estep_hard}.
			\ELSIF {stochastic assignment}
				\STATE $\Gamma \leftarrow \textsf{stochastic}(\Gamma)$ by Equation \eqref{clustering:Estep_stochastic}.
			\ENDIF
			\STATE \{M-step\}
			\FOR{$k=1:K$}
				\STATE $\pi_k^{(t+1)} = \sum_{n=1}^N \gamma_{nk} / N$.
				\STATE $\mu_k^{(t+1)} = \underset{\mu \in \bbS^p}{\argmin}  \sum_{n=1}^N \gamma_{nk} \cdot d^2 (x_n, \mu)$ by Algorithm \ref{code:lkd_mean}.
			\ENDFOR
			\IF {homogeneous concentration}
				\STATE compute $\lambda^{(t+1)}$ by Equation \eqref{clustering:rule_M_concentration_homo}.
			\ELSE
				\FOR {$k=1:K$}
					\STATE  compute $\lambda_k^{(t+1)}$ by Equation
					\eqref{clustering:rule_M_concentration_hetero}.
				\ENDFOR
			\ENDIF
		\UNTIL convergence.
	\end{algorithmic}
\end{algorithm}

We describe a few practical details of the EM algorithm. The algorithm is first initialized with standard $k$-means clustering \citep{macqueen_methods_1967, hornik_movmf_2014} since there exist a number of fast and efficient implementations of the standard $k$-means algorithm. Furthermore, an equivariant embedding on the sphere that preserves a large amount of geometric information is the identity map \citep{bhattacharya_nonparametric_2012-1}, which may justify the scheme by arguing that some degree of geometric information is preserved. Second, one popular choice for the termination criterion is to run iterations until the log likelihood \eqref{clustering:loglkd_naive} no longer increases. Although the quantity itself is an object of our interest, it may be computationally prohibitive for a large dataset since evaluating the log likelihood at each iteration requires to compute densities with updated parameters and no intermediate results can be re-used. One heuristic may terminate the algorithm if the distribution of latent membership variables does not change much. In other words, we suggest to use $\|\Gamma^{(t+1)} - \Gamma^{(t)}\|$ as a proxy for convergence since it only uses information already obtained while it implies that clustering results do not evolve.

Finally, we close this section with a remark on geodesic $k$-means algorithm as a special case of the mixture of SN distributions. In cluster analysis, the relationship between $k$-means algorithm and Gaussian mixture model has long been known \citep{bishop_pattern_2006}, which was also explored by the mixture of vMF distributions \citep{banerjee_clustering_2005} and the spherical $k$-means algorithm \citep{dhillon_concept_2001}. When homogeneous concentration parameter is used, a soft clustering matrix is given by 
\begin{equation}
\gamma_{nk} = \frac{\pi_k f_\textsf{SN}(x_n | \mu_k, \lambda)}{\sum_{j=1}^K \pi_j f_\textsf{SN}(x_n | \mu_j, \lambda)} = \frac{\pi_k \exp \lbrace -\frac{\lambda}{2} d^2 (x_n, \mu_k) \rbrace}{\sum_{j=1}^K \pi_j \exp \left\lbrace -\frac{\lambda}{2} d^2 (x_n, \mu_j) \right\rbrace}.
\end{equation}
which measures the probability to assign an $n$-th observation to $k$-th cluster. As $\lambda \rightarrow \infty$, the term where $d^2 (x_n, \mu_j)$ is smallest decays most slowly in the denominator. Consequently, $\gamma_{nk}$ approaches zero except for the index $j$ that attains the smallest distance to $x_n$. This scenario is equivalent to a hard assignment in the $k$-means clustering since the rule $\gamma_{nk} = 1$ if $k = \argmin_j d^2(x_n, \mu_j)$ matches the assignment step in the $k$-means clustering algorithm. This is also verified by re-writing the complete-data log likelihood in Equation \eqref{clustering:rule_M_equation} 
\begin{equation*}
\bbE_{\bfZ|\bfX,\bfTheta } \lbrack \log P(\bfX, \bfZ | \bfTheta) \rbrack \xrightarrow{\lambda \rightarrow \infty} -\frac{1}{2} \sum_{n=1}^N \sum_{k=1}^K \gamma_{nk} d^2(x_n, \mu_k) + \textrm{constant}
\end{equation*}
so that maximization of the expected complete-data log likelihood corresponds to minimize the cost function based on the Voronoi diagram from $k$-means algorithm in the limiting sense.

\section{Experiments}\label{sec:experiment}

We validate efficacy and performance of algorithms for parameter estimation and mixture modeling for the SN distributions with simulated and real data examples. For clustering, we compare the algorithms in Table \ref{table:alg_compare}. The concentration parameters of two mixture models with vMF and SN distributions are set as heterogeneous across all components. 
\begin{table}[ht]
	\begin{center}
		\begin{tabular}{|c|c|c|}
			\hline
			\multicolumn{2}{|c|}{algorithm}& description \\ \hline
			\textsc{kmeans} & \cite{macqueen_methods_1967} & $k$-means \\ \hline
			\textsc{spkmeans} & \cite{dhillon_concept_2001} & spherical $k$-means \\ \hline
			\textsc{vMF-soft}    & \multirow{2}{*}{\cite{banerjee_clustering_2005}} & von Mises-Fisher mixture with soft assignment   \\ \cline{1-1} \cline{3-3} 
			\textsc{vMF-hard}    &                   & von Mises-Fisher mixture  with hard assignment   \\ \hline
						\textsc{SN-soft}    & \multirow{2}{*}{Section \ref{sec:clustering}} & spherical normal mixture with soft assignment   \\ \cline{1-1} \cline{3-3} 
			\textsc{SN-hard}    &                   & spherical normal mixture  with hard assignment   \\ \hline
		\end{tabular}
	\end{center}
	\caption{Clustering algorithms for comparison.}
	\label{table:alg_compare}
\end{table}

When a true or desired clustering label is available, quality of an algorithm is assessed using clustering comparison indices including Rand index \citep{rand_objective_1971-2}, Jaccard index \citep{jaccard_distribution_1912-1}, and normalized mutual information \citep{strehl_cluster_2002}. All three indices have values in $[0,1]$ and share a characterization that if two clusterings are identical up to permutation or relabeling of assignments, all have a value of 1. For example, given three clusterings $S_0=(1,1,2,2)$, $S_1=(2,2,1,1)$ an $S_2=(1,1,3,3)$ of a set of 4 objects, any pairwise index among the three equals to 1.

\subsection{Simulation for Parameter Estimation}

We first compare performances of several algorithms that were proposed for the maximum likelihood estimation of location and concentration parameters. Given two fixed parameters $\mu_0 = (0,0,0,0,0,1)$ and $\lambda_0 = 10$ for the SN distribution on the 5-dimensional sphere $\bbS^5$, we assess how performance measures such as accuracy and wall-clock time for each run of an algorithm change as the number of observations $n$ varies from 25 to 475. We repeat each run 100 times and report empirical distributions of the performance measures. We used the stopping criterion $\epsilon = 10^{-8}$, which approximately equals to a square root of machine epsilon in double precision.

\begin{figure}[ht]
	\centering
	\includegraphics[width=0.9\linewidth]{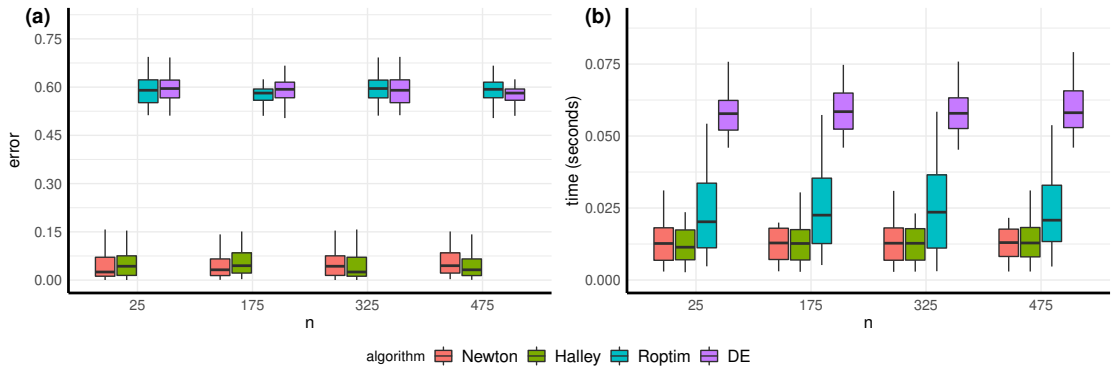}
	\caption{(a) Accuracy and (b) time elapsed for the concentration estimation problem.}
	\label{fig:simulation_concentration}
\end{figure}

First, we compare two step-size rules, line search and fixed size $\alpha=0.25$, for the location parameter estimation problem. Let $\hat{\mu}_{\text{MLE}} = \hat{\mu}_{\text{MLE}}(\bfX_{1:n})$ be a maximum likelihood estimate for a random sample of size $n$. The accuracy is represented by an error quantity $\|\hat{\mu}_{\text{MLE}}-\mu_0\|$. For varying $n$, the results are shown in Figure \ref{fig:simulation_concentration} that as $n$ gets larger, we obtain better estimates via either of the step-size rules while the computation cost increases as expected. One notable observation is that two rules are almost parallel in both estimation quality and computational cost. The latter phenomenon seems especially interesting. It is obvious that a single update with line search should outperform that with a fixed step-size rule. However, the line search requires repetitive evaluation of the cost function in a single iterate in that its accurate update compromises overall efficiency, leading to comparable computational costs in the end.
\begin{figure}[!ht]
	\centering
	\includegraphics[width=0.9\linewidth]{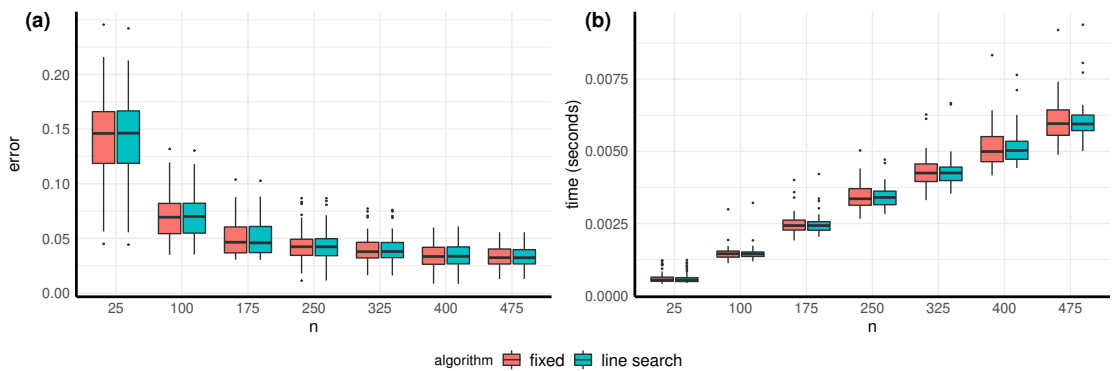}
	\caption{(a) Accuracy and (b) time elapsed for the location estimation problem.}
	\label{fig:simulation_location}
\end{figure}

We used the same approach for approximate Newton's and Halley's methods, which are compared along with two derivative-free optimization algorithms. One is \textsf{optimize}, a default optimization routine in \textsf{R} \citep{r_core_team_r_2021} which combines golden section search and successive parabolic interpolation \citep{brent_algorithms_2002}. We will denote the routine as \textsf{Roptim}. Next, \textsf{DE} stands for the celebrated differential evolution algorithm that approximates the global optimum for a real-valued function using a heuristic approach \citep{storn_differential_1997, mullen_deoptim_2011}. Accuracy of an estimate $\hat{\lambda}_{\text{MLE}} = \hat{\lambda}_{\text{MLE}}(\bfX_{1:n})$ is measured by relative error $|\hat{\lambda}_{\text{MLE}}-\lambda_0|/\lambda_0$ and the same level of stopping criterion $\epsilon = 10^{-8}$ is used for all testings. Performance measures are shown in Figure \ref{fig:simulation_concentration} where both approximate Newton's and Halley's methods consistently outperform the other two with smaller errors and shorter wall-clock time.

We also perform extensive simulations for performance comparison of the estimation algorithms with varying number of observations $n=50,100,150,200$. Table \ref{table:extended_location} summarizes performance measures in the location estimation problem for dimensions 
$p=5,10,20$ and concentrations $\lambda=5,10,50$. In every setting, we observed that line search and fixed step-size update rules perform almost tantamountly. Results from the concentration estimation problem are given by Table \ref{table:extended_concentration} for $p=5,10,20$ and $\lambda=1,5,10,20$, which show the same pattern as a fixed setting case that Newton's and Halley's methods show similar results while both have superior performance to the other two algorithms.

\subsection{Simulation for Clustering}

We consider two simulated examples, \textsf{small-mix} and \textsf{large-mix}, to validate clustering performance of the finite mixture of SN distributions, which are taken from \cite{banerjee_clustering_2005} and slightly modified. 

First, the \textsf{small-mix} example draws a random sample from a mixture of two SN components with parameters $(\mu_1,\lambda_1) = ([-0.251, -0.968], 10)$ and $(\mu_2,\lambda_2) = ([0.399, 0.917],2)$ on $\bbS^1 \subset \bbR^2$. For each component, 100 observations are randomly drawn. 
\begin{figure}[ht]
	\centering
	\includegraphics[width=0.99\linewidth]{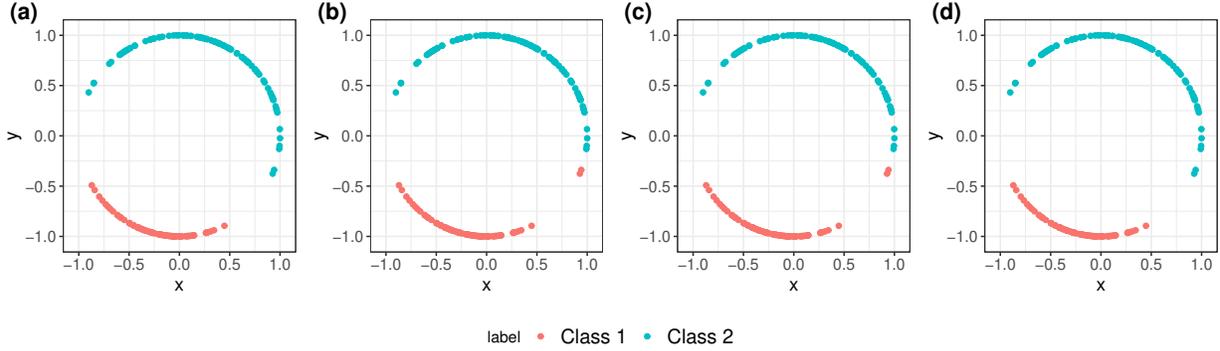}
	\caption{Visualization of the \textsf{small-mix} example of (a) the  data and clustering results from (b) \textsc{spkmeans}, (c) \textsc{vMF-soft}, and (d) \textsc{SN-soft} algorithms with $K=2$. }
	\label{fig:small_mix}
\end{figure}
A random sample and some clustering results are given in Figure \ref{fig:small_mix} Both \textsc{spkmeans} and \textsc{vMF-soft} showed a slight misclassification for the bottom-most observations of the 2nd class in contrast to \textsc{SN-soft}. The same test was repeated 10 times and average clustering quality indices are reported in Table \ref{table:small_mix} with different numbers of clusters $K=2,3$ and $4$. When $K=2$, both \textsc{SN} and \textsc{vMF} mixtures performed well while the former had slightly better results. When $K>2$, \textsc{SN} mixtures had less accurate results than \textsc{vMF} mixtures, which is reasonable and rather convincing to claim fitness of \textsc{SN} mixtures since the true model consists of \textsc{SN} components and inflated quality indices of \textsc{vMF} mixtures may potentially confuse the decision given such results. It is not surprising that \textsc{spkmeans} showed compatible result as a limit of  \textsc{vMF} mixture model. In the \textsf{small-mix} example, the data has two linearly separable components so that the $k$-means algorithm performed competitively.

\begin{table}[ht]
	\begin{center}
			\begin{tabular}{|c|c|c|c|c|c|c|c|c|c|}
			\hline
			\multirow{2}{*}{} & \multicolumn{3}{c|}{Rand} & \multicolumn{3}{c|}{Jaccard} & \multicolumn{3}{c|}{NMI} \\ \cline{2-10} 
			& $K=2$     & $K=3$    & $K=4$    & $K=2$      & $K=3$     & $K=4$     & $K=2$   & $K=3$   & $K=4$    \\ \hline
\textsc{kmeans} & 0.9613 & 0.7250 & 0.5709 & 0.9802 & 0.8614 & 0.7852 & 0.9327 & 0.7633 & 0.7047 \\ \hline
\textsc{spkmeans} & 0.9408 & 0.7454 & 0.6264 & 0.9694 & 0.8723 & 0.8131 & 0.9025 & 0.7969 & 0.7231 \\ \hline
\textsc{vMF-soft} & 0.9820 & 0.8152 & 0.6883 & 0.9960 & 0.9075 & 0.8443 & 0.9738 & 0.8356 & 0.7558 \\ \hline
\textsc{vMF-hard} & 0.9820 & 0.9284 & 0.8422 & 0.9960 & 0.9643 & 0.9212 & 0.9738 & 0.9144 & 0.8417 \\ \hline
\textsc{SN-soft} & 0.9920 & 0.8025 & 0.6834 & 0.9960 & 0.9011 & 0.8416 & 0.9838 & 0.8257 & 0.7490 \\ \hline
\textsc{SN-hard} & 0.9881 & 0.7741 & 0.7054 & 0.9940 & 0.8866 & 0.8526 & 0.9767 & 0.8099 & 0.7600 \\  \hline
		\end{tabular}
	\end{center}
	\caption{Average clustering quality indices from 10 runs with different numbers of clusters for the \textsf{small-mix} example.}
	\label{table:small_mix}
\end{table}

The \textsf{large-mix} example considers a mixture density on $\bbS^3$ that is composed of 3 SN distributions with location parameters that are drawn randomly to reside in different quadrants, concentration parameters  $(\lambda_1,\lambda_2,\lambda_3)=(40,20,60)$. Component weights are chosen as follows. Draw $u_i \sim U(9,11)$ for $i=1,2,3$ and set $\pi_i = u_i / \sum_{i=1}^3 u_i$, which assigns almost equal weights for individual components. We draw a random sample of 3000 observations according to the specified mixture model and the results for average of 10 runs are reported in Table \ref{table:large_mix}. 
\begin{table}[ht]
	\begin{center}
		\begin{tabular}{|c|c|c|c|c|c|c|c|c|c|}
			\hline
			\multirow{2}{*}{} & \multicolumn{3}{c|}{Rand} & \multicolumn{3}{c|}{Jaccard} & \multicolumn{3}{c|}{NMI} \\ \cline{2-10} 
			& $K=2$     & $K=3$    & $K=4$    & $K=2$      & $K=3$     & $K=4$     & $K=2$   & $K=3$   & $K=4$    \\ \hline
			\textsc{kmeans} & 0.6413 & 0.5051 & 0.7736 & 0.8103 & 0.7383 & 0.9226 & 0.7806 & 0.6607 & 0.8764 \\ \hline
			\textsc{spkmeans} & 0.6413 & 0.9844 & 0.7744 & 0.8103 & 0.9947 & 0.9229 & 0.7806 & 0.9764 & 0.8766 \\ \hline
			\textsc{vMF-soft} & 0.6413 & 0.9856 & 0.9806 & 0.8103 & 0.9951 & 0.9934 & 0.7806 & 0.9781 & 0.9690 \\ \hline
			\textsc{vMF-hard} & 0.6413 & 0.9856 & 0.9819 & 0.8103 & 0.9951 & 0.9938 & 0.7806 & 0.9781 & 0.9733 \\ \hline
			\textsc{SN-soft} & 0.6413 & 0.9856 & 0.7902 & 0.8103 & 0.9951 & 0.8283 & 0.7806 & 0.9781 & 0.8825 \\ \hline
			\textsc{SN-hard} & 0.6413 & 0.9856 & 0.8833 & 0.8103 & 0.9951 & 0.8943 & 0.7806 & 0.9781 & 0.8750 \\ \hline
		\end{tabular}
	\end{center}
	\caption{Average clustering quality indices from 10 runs with different numbers of clusters for the \textsf{large-max} example.}
	\label{table:large_mix}
\end{table}
Since we are using large concentration parameters with mutually distant locations, it is expected that observations per class should have little overlapping support and any good model should be able to detect three distinct clusters. In other words, a quality index should be low for $K=2,4$ and high for $K=3$. Except for the $k$-means, all algorithms perform exceptionally well when $K=3$. However, the desired pattern is only observed for \textsc{SN} mixtures and \textsc{spkmeans} though the latter sometimes does not decrease as much as the former. It is worth to mention that  \textsc{vMF} mixtures even show higher quality indices when $K=4$.

\subsection{Real Data Analysis}

We now validate clustering performance of the proposed mixture model with SN distributions on two real data - \textsf{household} and \textsf{Classic3}.

The \textsf{household} data is part of a survey data of household expenditures on commodity groups such as housing, goods, service, and food among 20 single males and females \citep{hothorn_handbook_2014}. We follow the convention of \cite{hornik_movmf_2014} to focus on relative portion of total expenditures for housing, service, and food categories only. Each individual's expenditure profile in $\bbR^3$ is projected onto $\bbS^2$ by $L_2$ normalization $x \leftarrow x/\|x\|$. As a preliminary step, we fitted the SN distribution to expenditures of males and females separately whose maximum likelihood estimates are shown in Table \ref{table:household_parameters}. We note that the females' expenditures have a larger concentration by $\hat{\lambda}_{\text{MLE}}=95.743$ while the males show much dispersed pattern as indicated by $\hat{\lambda}_{\text{MLE}}=19.638$. 
\begin{table}[ht]
	\begin{center}
		\begin{tabular}{|c|c|c|}
			\hline 
			gender & location & concentration \\ \hline
			female & $(0.954, 0.266, 0.135)$ & $95.743$ \\ \hline
			male & $(0.643, 0.407, 0.648)$ & $19.638$ \\ \hline
		\end{tabular}
	\end{center}
	\caption{Maximum likelihood estimates of the spherical normal distributions on projected commodity expenditures from the \textsf{household} data by gender.}
	\label{table:household_parameters}
\end{table}

\begin{figure}[ht]
	\centering
	\includegraphics[width=0.9\linewidth]{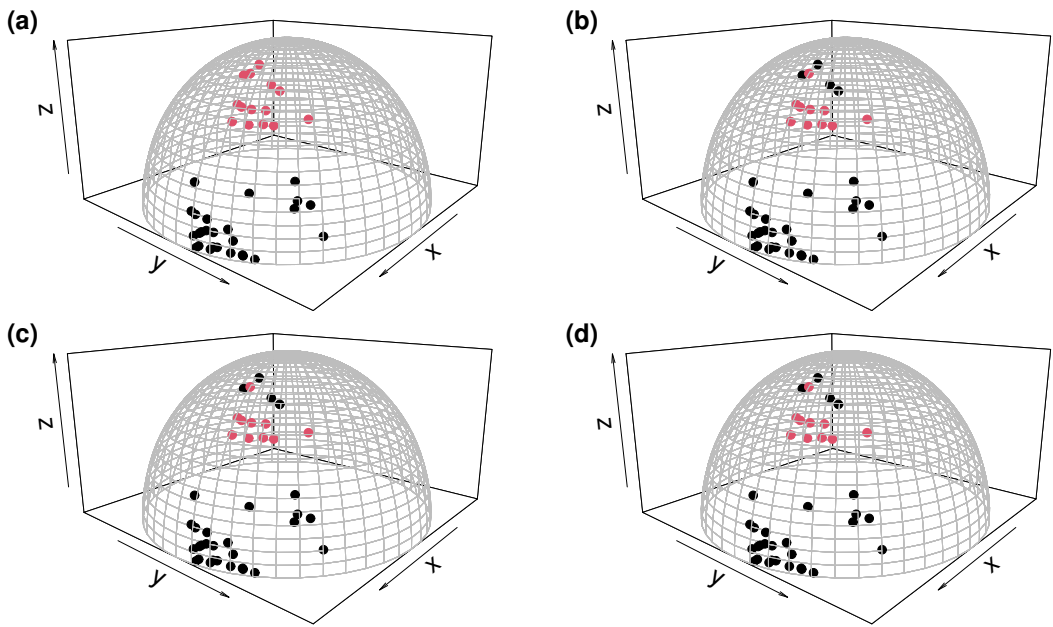}
	\caption{Visualization of the preprocessed \textsf{household} data colored by (a) gender and clusterings from (b) \textsc{kmeans}, (c) \textsc{spkmeans}, and (d) \textsc{SN-soft} algorithms for $K=2$.}
	\label{fig:household_visualization}
\end{figure}

We first use the gender as ground-truth clustering and perform clustering where the results are shown in Figure \ref{fig:household_visualization}. All three clusterings of \textsc{kmeans}, \textsc{spkmeans}, and \textsc{SN-soft} algorithms misclassified some females. We can observe this phenomenon in a more extensive experiment as before. Table \ref{table:household_compare} shows that only the \textsc{vMF} mixtures recover two gender-defined clusters while other algorithms show higher degree of accuracy for $K=3$ consistently. 

\begin{table}[ht]
	\begin{center}
		\begin{tabular}{|c|c|c|c|c|c|c|c|c|c|}
			\hline
			\multirow{2}{*}{} & \multicolumn{3}{c|}{Rand} & \multicolumn{3}{c|}{Jaccard} & \multicolumn{3}{c|}{NMI} \\ \cline{2-10} 
			& $K=2$     & $K=3$    & $K=4$    & $K=2$      & $K=3$     & $K=4$     & $K=2$   & $K=3$   & $K=4$    \\ \hline
\textsc{kmeans} & 0.5920 & 0.7789 & 0.5363 & 0.7385 & 0.8923 & 0.7705 & 0.5105 & 0.8331 & 0.6546 \\ \hline
\textsc{spkmeans} & 0.5920 & 0.7789 & 0.5363 & 0.7385 & 0.8923 & 0.7705 & 0.5105 & 0.8331 & 0.6546 \\ \hline
\textsc{vMF-soft} & 0.9025 & 0.7275 & 0.6450 & 0.9500 & 0.8603 & 0.8179 & 0.8558 & 0.7244 & 0.6645 \\ \hline
\textsc{vMF-hard} & 0.9025 & 0.7275 & 0.6921 & 0.9500 & 0.8603 & 0.8500 & 0.8558 & 0.7244 & 0.7663 \\ \hline
\textsc{SN-soft} & 0.5920 & 0.7275 & 0.5363 & 0.7385 & 0.8603 & 0.7705 & 0.5105 & 0.7244 & 0.6546 \\ \hline
\textsc{SN-hard} & 0.5920 & 0.7789 & 0.5363 & 0.7385 & 0.8923 & 0.7705 & 0.5105 & 0.8331 & 0.6546 \\ \hline
		\end{tabular}
	\end{center}
	\caption{Clustering quality indices with different numbers of clusters for the \textsf{household} data.}
	\label{table:household_compare}
\end{table}

However, a careful scrutiny of the data allows us  to argue that the seemingly counter-intuitive observation is not much invalid. When fitted with the SN distribution, the male group turned out to have a smaller concentration. In other words, expenditures of males are highly dispersed which is also shown in the Figure \ref{fig:household_visualization} where there are several observations lying apart from a dense set near the equator. This makes an assertion highly plausible that the male group has two ingrained subgroups. 

\begin{figure}[ht]
	\centering
	\includegraphics[width=0.99\linewidth]{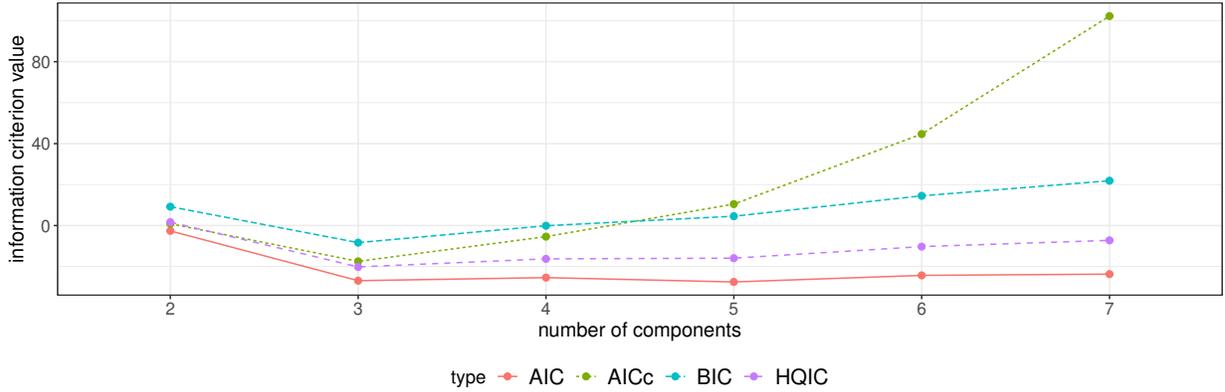}
	\caption{Several information criteria for the mixture of spherical normal distributions with varying number of clusters from $K=2$ to $K=7$.}
	\label{fig:household_criterion}
\end{figure}

In order to account for our statement, we report several information criteria of fitted \textsc{SN-soft} models with varying number of clusters, including Akaike information criterion \citep[AIC;][]{akaike_new_1974}, corrected AIC \citep[AICc;][]{cavanaugh_unifying_1997}, Bayesian information criterion \citep[BIC;][]{schwarz_estimating_1978-1}, and Hannan-Quinn information criterion \citep[HQIC;][]{hannan_determination_1979-1}. For a mixture with $K$ components on $\bbS^p$, the number of parameters with heterogeneous concentration is 
\begin{equation*}
k^* = p K + K + (K - 1) = (p+2)K-1
\end{equation*}
so that we have 
\begin{align*}
\text{AIC} &= -2\hat{L} + 2k^*\\
\text{AICc} &= \text{AIC} + \frac{2k^*(k^*+1)}{(N-k^*-1)} \\
\text{BIC} &= -2\hat{L} + k^* \log N \\
\text{HQIC} &= -2\hat{L} + 2 k^* \log(\log N)
\end{align*}
where $\hat{L}$ is the log-likelihood of an estimated SN mixture and $N$ denotes the sample size. Information criteria values for the \textsf{household} data are given in Figure \ref{fig:household_criterion}  which shows all but AIC report to have the minimal information criterion value at $K=3$ to support our proposition.

\begin{figure}[ht]
	\centering
	\includegraphics[width=0.99\linewidth]{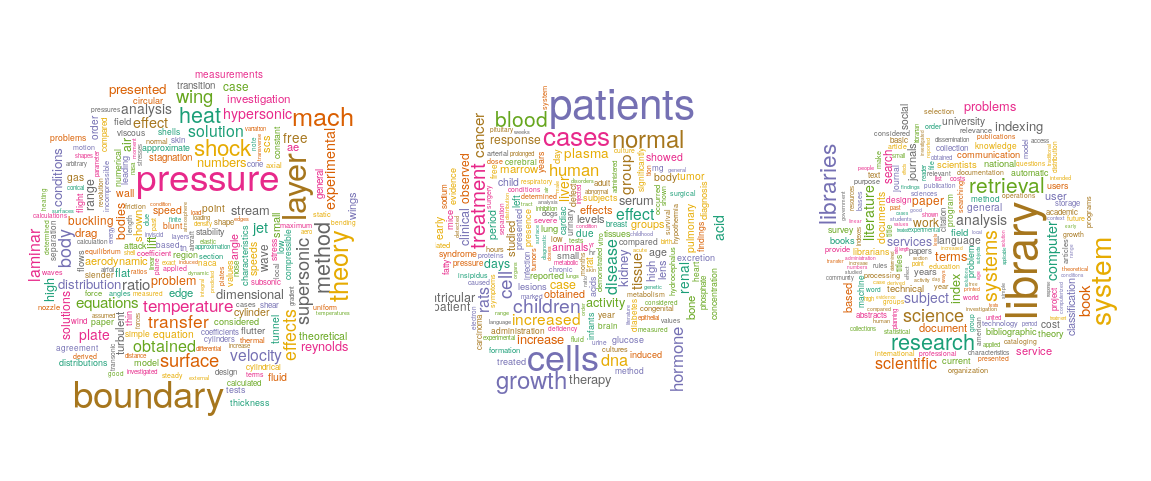}
	\caption{Word clouds of the most frequent terms according to the clustering generated by \textsc{SN-soft} with $K=3$ that correspond to the field of aeronautical system (left), medicine (middle), and information retrieval (right).}
	\label{fig:classic3}
\end{figure}

The second example is \textsf{Classic3} corpus extracted from a collection of documents from three distinct academic domains \citep{dhillon_information-theoretic_2003}. The data set contains a total of 3898 documents, among which 1400, 1033, and 1460 documents are drawn from studies of aeronautical system, medicine, and information retrieval, respectively. We trim the terms to have 4303 words left where each word appears in at least 8 and no more than 578 documents.
\begin{table}[ht]
	\begin{center}
		\begin{tabular}{|c|c|c|c|c|c|c|c|c|c|}
			\hline
			\multirow{2}{*}{} & \multicolumn{3}{c|}{Rand} & \multicolumn{3}{c|}{Jaccard} & \multicolumn{3}{c|}{NMI} \\ \cline{2-10} 
			& $K=2$     & $K=3$    & $K=4$    & $K=2$      & $K=3$     & $K=4$     & $K=2$   & $K=3$   & $K=4$    \\ \hline
\textsc{kmeans} & 0.6760 & 0.8645 & 0.8157 & 0.5935 & 0.8008 & 0.7632 & 0.6760 & 0.8645 & 0.7157 \\ 
\textsc{spkmeans} & 0.6832 & 0.9614 & 0.8029 & 0.6010 & 0.8929 & 0.7296 & 0.6832 & 0.9614 & 0.8029 \\ 
\textsc{vMF-soft} & 0.6992 & 0.9534 & 0.8218 & 0.5255 & 0.9271 & 0.7806 & 0.6992 & 0.9534 & 0.7218 \\ 
\textsc{vMF-hard} & 0.6994 & 0.9217 & 0.7268 & 0.5255 & 0.9249 & 0.7616 & 0.6994 & 0.9517 & 0.7268 \\ 
\textsc{SN-soft} & 0.6537 & 0.9645 & 0.8157 & 0.5638 & 0.9108 & 0.7632 & 0.6537 & 0.9645 & 0.6157 \\ 
\textsc{SN-hard} & 0.7537 & 0.9218 & 0.8121 & 0.5638 & 0.9050 & 0.7632 & 0.6537 & 0.9218 & 0.7078 \\ \hline
		\end{tabular}
	\end{center}
	\caption{Clustering quality indices with different numbers of clusters for the \textsf{Classic3} data.}
	\label{table:classic3}
\end{table}
Table \ref{table:classic3} summarizes clustering performance of the algorithms, all of which show best results when $K=3$. However, the standard $k$-means algorithm does not match those of competing algorithms in all quality indices. Based on a clustering result of \textsc{SN-soft}, the relative frequency of terms appearance per class is demonstrated using word clouds in Figure \ref{fig:classic3} where class-specific terms make noticeable differences for grouping the documents. 

\section{Conclusion}\label{sec:conclusion}

In this work, we discussed numerical schemes for maximum likelihood estimation of the SN distribution on the unit hypersphere, which was recently proposed by \cite{hauberg_directional_2018-1} as an intrinsic counterpart to the vMF distribution. We showed equivalence of the location estimation problem to the \Frechet mean computation on a Riemannian manifold. The algorithms for concentration parameter estimation were proposed upon the low-order Householder method combined with finite difference approximation of the derivatives and showed superb performance in terms of both accuracy and efficiency than naive black-box optimization methods. Furthermore, we considered parameter estimation problems in a more general setting with a set of weighted observations, which led to another contribution of ours to elaborate updating rules of the finite mixture model using an expectation-maximization algorithm.

We believe our contributions provide opportunities for immediate uses and interesting future studies in directional statistics. For example, most of traditional hypothesis testing routines on the unit hypersphere are based on parameter estimates of vMF distributions  \citep{mardia_directional_2000}, which can be directly substituted by those obtained from our proposal. In clustering, our proposal can benefit modification of well-studied algorithms built upon vMF distribution. In \cite{gopal_2014_MisesfisherClusteringModels}, Bayesian framework was proposed for several contexts including  standard, hierarchical, and temporal mixtures. Since there is no conjugacy available for SN distribution, it is of significant importance to devise efficient numerical routines for a variety of SN mixtures, which we expect a lot of works left for future studies.

\section*{Appendix}
\begin{center}
	\begin{longtable}{|L|L|L|L|L|L|L|}
		\caption{Accuracy and elapsed time for the location parameter estimation.}
		\label{table:extended_location}\\
		\hline
		\multirow{2}{*}{dimension}          & \multirow{2}{*}{$\lambda$} & \multirow{2}{*}{$n$} & \multicolumn{2}{c|}{line search} & \multicolumn{2}{c|}{fixed step-size} \\ \cline{4-7} 
		&                     &    & accuracy & time & accuracy & time \\ \hhline{|=|=|=|=|=|=|=|}
		\multirow[c]{12}{*}{$p=5$} 
		& \multirow{4}{*}{5} & 50   & 0.15510 & 0.00054 & 0.15473 & 0.00054\\ \cline{3-7} 
		&                    & 100  & 0.10797 & 0.00081 & 0.10788 & 0.00082\\ \cline{3-7} 
		&                    & 150  & 0.06682 & 0.00188 & 0.06668 & 0.00190 \\ \cline{3-7} 
		&                    & 200  & 0.04935 & 0.00372 & 0.04933 & 0.00371 \\ \cline{2-7} 
		& \multirow{4}{*}{10} & 50  & 0.09602 & 0.00081 & 0.09628 & 0.00081 \\ \cline{3-7} 
		&                     & 100 & 0.06886 & 0.00142 & 0.06908 & 0.00141 \\ \cline{3-7} 
		&                     & 150 & 0.04503 & 0.00315 & 0.04503 & 0.00314 \\ \cline{3-7} 
		&                     & 200 & 0.03400 & 0.00621 & 0.03405 & 0.00617  \\ \cline{2-7}
		& \multirow{4}{*}{50} & 50  & 0.04392 & 0.00104 & 0.04388 & 0.00104 \\ \cline{3-7} 
		&                     & 100 & 0.03035 & 0.00197 & 0.03043 & 0.00196\\ \cline{3-7} 
		&                     & 150 & 0.02114 & 0.00487 & 0.02111 & 0.00491 \\ \cline{3-7} 
		&                     & 200 & 0.01278 & 0.00967 & 0.01276 & 0.00982 \\ \hline 
		\multirow{12}{*}{$p=10$} 
		& \multirow{4}{*}{5} & 50   & 0.24234 & 0.00032 & 0.24221 & 0.00032 \\ \cline{3-7} 
		&                    & 100  & 0.17989 & 0.00053 & 0.18004 & 0.00053\\ \cline{3-7} 
		&                    & 150  & 0.11274 & 0.00102 & 0.11277 & 0.00103 \\ \cline{3-7} 
		&                    & 200  & 0.08021 & 0.00219 & 0.08004 & 0.00220 \\ \cline{2-7} 
		& \multirow{4}{*}{10} & 50  & 0.15318 & 0.00059 & 0.15268 & 0.00059 \\ \cline{3-7} 
		&                     & 100 & 0.11267 & 0.00079 & 0.11269 & 0.00079 \\ \cline{3-7} 
		&                     & 150 & 0.06913 & 0.00175 & 0.06909 & 0.00174 \\ \cline{3-7} 
		&                     & 200 & 0.04608 & 0.00340 & 0.04601 & 0.00342  \\ \cline{2-7}
		& \multirow{4}{*}{50} & 50  & 0.06176 & 0.00113 & 0.06170 & 0.00113\\ \cline{3-7} 
		&                     & 100 & 0.04424 & 0.00214 & 0.04427 & 0.00218\\ \cline{3-7} 
		&                     & 150 & 0.02805 & 0.00530 & 0.02804 & 0.00532\\ \cline{3-7} 
		&                     & 200 & 0.02136 & 0.01076 & 0.02136 & 0.01078\\ \hline 	
		\multirow{12}{*}{$p=20$} 
		& \multirow{4}{*}{5} & 50   & 0.40135 & 0.00057 & 0.40214 & 0.00057   \\ \cline{3-7} 
		&                    & 100  & 0.30380 & 0.00092 & 0.30349 & 0.00092  \\ \cline{3-7} 
		&                    & 150  & 0.18930 & 0.00246 & 0.18883 & 0.00249  \\ \cline{3-7} 
		&                    & 200  & 0.13773 & 0.00424 & 0.13760 & 0.00421  \\ \cline{2-7} 
		& \multirow{4}{*}{50} & 50  & 0.25568 & 0.00056 & 0.25572 & 0.00056  \\ \cline{3-7} 
		&                     & 100 & 0.18253 & 0.00101 & 0.18225 & 0.00101 \\ \cline{3-7} 
		&                     & 150 & 0.11150 & 0.00239 & 0.11143 & 0.00235 \\ \cline{3-7} 
		&                     & 200 & 0.08036 & 0.00409 & 0.08024 & 0.00407  \\ \cline{2-7}
		& \multirow{4}{*}{50} & 50  & 0.09484 & 0.00215 & 0.09468 & 0.00216 \\ \cline{3-7} 
		&                     & 100 & 0.07044 & 0.00405 & 0.07051 & 0.00409 \\ \cline{3-7} 
		&                     & 150 & 0.04325 & 0.01073 & 0.04329 & 0.01074 \\ \cline{3-7} 
		&                     & 200 & 0.02985 & 0.01750 & 0.02986 & 0.01759  \\ \hline 			
	\end{longtable}
\end{center}

\begin{center}
	\begin{longtable}{|c|c|c|c|c|c|c|c|c|c|c|}
		\caption{Accuracy and elapsed time for the concentration parameter estimation.} \label{table:extended_concentration}\\
		\hline
		\multirow{2}{*}{$\lambda$} & \multirow{2}{*}{$p$} & \multirow{2}{*}{$n$} & \multicolumn{4}{c|}{accuracy} & \multicolumn{4}{c|}{time (seconds)} \\ \cline{4-11} 
		&                     &    & Newton & Halley & \textsf{Roptim} & \textsf{DE} & Newton & Halley & \textsf{Roptim} & \textsf{DE} \\ \hhline{|=|=|=|=|=|=|=|=|=|=|=|}
		\multirow{12}{*}{$1$}   
		& \multirow{4}{*}{5}  & 50  & 0.23140 & 0.23140 & 0.23140 & 0.23140 & 0.00319 & 0.00338 & 0.00362 & 0.09734\\ \cline{3-11} 
		&                     & 100 & 0.17456 & 0.17456 & 0.17456 & 0.17456 & 0.00490 & 0.00507 & 0.00583 & 0.09955\\ \cline{3-11} 
		&                     & 150 & 0.09902 & 0.09902 & 0.09902 & 0.09902 & 0.00934 & 0.00956 & 0.01329 & 0.10369\\ \cline{3-11} 
		&                     & 200 & 0.06832 & 0.06832 & 0.06831 & 0.06832 & 0.01807 & 0.01778 & 0.02647 & 0.11374\\ \cline{2-11} 
		& \multirow{4}{*}{10} & 50  & 0.43744 & 0.43744 & 0.43744 & 0.43744 & 0.00356 & 0.00375 & 0.00374 & 0.10118\\ \cline{3-11} 
		&                     & 100 & 0.28210 & 0.28210 & 0.28210 & 0.28210 & 0.00547 & 0.00570 & 0.00657 & 0.10579\\ \cline{3-11} 
		&                     & 150 & 0.13216 & 0.13216 & 0.13216 & 0.13216 & 0.01039 & 0.01053 & 0.01388 & 0.10410\\ \cline{3-11} 
		&                     & 200 & 0.11031 & 0.11031 & 0.11031 & 0.11031 & 0.01970 & 0.02076 & 0.02879 & 0.11676\\ \cline{2-11}
		& \multirow{4}{*}{20} & 50  & 1.31015 & 1.31015 & 1.31015 & 1.31015 & 0.00508 & 0.00499 & 0.00548 & 0.12191\\ \cline{3-11} 
		&                     & 100 & 0.71212 & 0.71212 & 0.71212 & 0.71212 & 0.00881 & 0.00898 & 0.01025 & 0.13198\\ \cline{3-11} 
		&                     & 150 & 0.34126 & 0.34126 & 0.34127 & 0.34126 & 0.01889 & 0.01908 & 0.02605 & 0.14103\\ \cline{3-11} 
		&                     & 200 & 0.20973 & 0.20973 & 0.20973 & 0.20973 & 0.03835 & 0.03744 & 0.05134 & 0.15333\\ \hline
		\multirow{12}{*}{$5$}   
		& \multirow{4}{*}{5}  &50  & 0.11958 & 0.11958 & 0.09638 & 0.09638 & 0.00375 & 0.00390 & 0.00401 & 0.10405 \\   \cline{3-11} 
		&                     &100 & 0.07568 & 0.07568 & 0.07238 & 0.07238 & 0.00567 & 0.00544 & 0.00611 & 0.10007 \\   \cline{3-11} 
		&                     &150 & 0.04168 & 0.04168 & 0.04168 & 0.04168 & 0.01054 & 0.01039 & 0.01425 & 0.10640 \\   \cline{3-11} 
		&                     &200 & 0.03315 & 0.03315 & 0.03315 & 0.03315 & 0.01940 & 0.01910 & 0.03280 & 0.11397 \\   \cline{2-11} 
		& \multirow{4}{*}{10} &50  & 0.09433 & 0.09433 & 0.09140 & 0.09140 & 0.00346 & 0.00344 & 0.00355 & 0.09485 \\  \cline{3-11} 
		&                     &100 & 0.05784 & 0.05784 & 0.06353 & 0.06353 & 0.00501 & 0.00498 & 0.00609 & 0.09748 \\   \cline{3-11} 
		&                     &150 & 0.03738 & 0.03738 & 0.03567 & 0.03567 & 0.01026 & 0.00990 & 0.01328 & 0.10234 \\   \cline{3-11} 
		&                     &200 & 0.02159 & 0.02159 & 0.02037 & 0.02037 & 0.01763 & 0.01745 & 0.02407 & 0.11142 \\   \cline{2-11}
		& \multirow{4}{*}{20} &50  & 0.14114 & 0.14114 & 0.29826 & 0.29824 & 0.00442 & 0.00441 & 0.00545 & 0.11437 \\  \cline{3-11} 
		&                     &100 & 0.08231 & 0.08231 & 0.27433 & 0.27432 & 0.00614 & 0.00637 & 0.00755 & 0.11564 \\   \cline{3-11} 
		&                     &150 & 0.04570 & 0.04570 & 0.26554 & 0.26553 & 0.01210 & 0.01144 & 0.01596 & 0.12392 \\  \cline{3-11} 
		&                     &200 & 0.03674 & 0.03674 & 0.26211 & 0.26210 & 0.02134 & 0.02098 & 0.02928 & 0.13025 \\   \hline
		\multirow{12}{*}{$10$}   
		& \multirow{4}{*}{5}  &50  & 0.09592 & 0.09592 & 0.58344 & 0.58343 & 0.00485 & 0.00448 & 0.00539 & 0.10564 \\  \cline{3-11} 
		&                     &100 & 0.07381 & 0.07381 & 0.58345 & 0.58344 & 0.00713 & 0.00737 & 0.01006 & 0.11539 \\  \cline{3-11} 
		&                     &150 & 0.04226 & 0.04226 & 0.58779 & 0.58779 & 0.01286 & 0.01253 & 0.01727 & 0.11279 \\  \cline{3-11} 
		&                     &200 & 0.03060 & 0.03060 & 0.58378 & 0.58377 & 0.02209 & 0.02221 & 0.03064 & 0.11539 \\  \cline{2-11} 
		& \multirow{4}{*}{10} &50  & 0.07377 & 0.07377 & 0.64883 & 0.64882 & 0.00394 & 0.00386 & 0.00446 & 0.09885 \\ \cline{3-11} 
		&                     &100 & 0.05496 & 0.05496 & 0.63979 & 0.63979 & 0.00536 & 0.00553 & 0.00696 & 0.10279 \\  \cline{3-11} 
		&                     &150 & 0.02660 & 0.02660 & 0.63858 & 0.63857 & 0.01078 & 0.01101 & 0.01440 & 0.10001 \\  \cline{3-11} 
		&                     &200 & 0.02046 & 0.02046 & 0.63544 & 0.63543 & 0.01922 & 0.01911 & 0.02705 & 0.11107 \\  \cline{2-11}
		& \multirow{4}{*}{20} &50  & 0.09019 & 0.09019 & 0.73180 & 0.73180 & 0.00464 & 0.00454 & 0.00497 & 0.11977 \\ \cline{3-11} 
		&                     &100 & 0.05920 & 0.05920 & 0.72993 & 0.72992 & 0.00669 & 0.00620 & 0.00750 & 0.11804 \\  \cline{3-11} 
		&                     &150 & 0.02901 & 0.02901 & 0.72192 & 0.72191 & 0.01208 & 0.01198 & 0.01789 & 0.12330 \\  \cline{3-11} 
		&                     &200 & 0.02233 & 0.02233 & 0.72563 & 0.72563 & 0.02196 & 0.02112 & 0.03136 & 0.13584 \\  \hline
		\multirow{12}{*}{$20$}   
		& \multirow{4}{*}{5}  &50  & 0.08496 & 0.08496 & 0.88850 & 0.88849 & 0.00471 & 0.00468 & 0.00545 & 0.10104 \\   \cline{3-11} 
		&                     &100 & 0.05488 & 0.05488 & 0.89197 & 0.89196 & 0.00722 & 0.00678 & 0.00867 & 0.10400 \\   \cline{3-11} 
		&                     &150 & 0.03153 & 0.03153 & 0.88959 & 0.88959 & 0.01426 & 0.01432 & 0.01933 & 0.10787 \\   \cline{3-11} 
		&                     &200 & 0.01993 & 0.01993 & 0.88957 & 0.88957 & 0.02565 & 0.02654 & 0.03722 & 0.12373 \\   \cline{2-11} 
		& \multirow{4}{*}{10} &50  & 0.06118 & 0.06118 & 0.90044 & 0.90043 & 0.00477 & 0.00462 & 0.00488 & 0.10417 \\  \cline{3-11} 
		&                     &100 & 0.05391 & 0.05391 & 0.89876 & 0.89876 & 0.00651 & 0.00690 & 0.00782 & 0.10122 \\   \cline{3-11} 
		&                     &150 & 0.02796 & 0.02796 & 0.89777 & 0.89776 & 0.01194 & 0.01212 & 0.01724 & 0.10729 \\  \cline{3-11} 
		&                     &200 & 0.02032 & 0.02032 & 0.89553 & 0.89553 & 0.02275 & 0.02301 & 0.03272 & 0.11606 \\   \cline{2-11}
		& \multirow{4}{*}{20} &50  & 0.06101 & 0.06101 & 0.91314 & 0.91313 & 0.00584 & 0.00537 & 0.00622 & 0.12590 \\  \cline{3-11} 
		&                     &100 & 0.03630 & 0.03630 & 0.91254 & 0.91253 & 0.00781 & 0.00785 & 0.00922 & 0.12984 \\   \cline{3-11} 
		&                     &150 & 0.02759 & 0.02759 & 0.91001 & 0.91001 & 0.01702 & 0.01653 & 0.02343 & 0.15604 \\   \cline{3-11} 
		&                     &200 & 0.01559 & 0.01559 & 0.90714 & 0.90714 & 0.03008 & 0.03056 & 0.04372 & 0.17168 \\   \hline
	\end{longtable}
\end{center}

\bibliographystyle{dcu}
\bibliography{reference}

\end{document}